\documentclass[12pt]{article}
\usepackage{amsmath,amssymb,bbold,epsfig}
\usepackage{amsfonts}
\usepackage{slashed}
\usepackage{graphicx,color}
\usepackage[sort&compress,comma,square]{natbib}
\usepackage{bbm}
\usepackage{ctable,longtable}
\usepackage[hypertex]{hyperref}

\newcommand{\re}[1]{\mathop{{\rm Re}\left[#1\right]}}
\newcommand{\im}[1]{\mathop{{\rm Im}\left[#1\right]}}

\newcommand\al{\alpha}
\newcommand\be{\beta}

\newcommand\ep{\epsilon}

\newcommand\beq{\begin{equation}}
\newcommand\eeq{\end{equation}}
\newcommand\bea{\begin{eqnarray}}
\newcommand\eea{\end{eqnarray}}
\newcommand\bi{\begin{itemize}}
\newcommand\ei{\end{itemize}}
\newcommand\ben{\begin{enumerate}}
\newcommand\een{\end{enumerate}}

\def\dfrac#1#2{{\displaystyle\frac{#1}{#2}}}
\newcommand\schd{Schr$\ddot{\rm o}$dinger}

\newcommand\sch{Schr$\ddot{\rm o}$dinger~}

\newcommand\poi{Poincar$\acute{\rm e}$~}

\newcommand\bloch{Bl$\ddot{\rm o}$ch~}

\newcommand\bp{{\bf p}}
\newcommand\bq{{\bf q}}
\def\ket#1{| \,#1\, \rangle}
\def\bra#1{\langle \,#1\, |}
\def\scx#1#2{\langle \,#1\, |\, #2\, \rangle}

\def\me#1#2#3{\langle \,#2\, |\,#1\,|\, #3\,\rangle}
\newif\ifboo \boofalse

\renewcommand{\vec}[1]{{\mathbf{#1}}}

\begin{document}

\title{\vspace{-2cm}
\hfill {\small }\\
\vglue 0.8cm \hfill {\small arXiv: 0907.0562v2 [hep-ph]} \vskip
0.5cm \Large \bf Geometric imprint of CP violation in two flavor
neutrino oscillations}
\author{
{Poonam Mehta$$\thanks{email: \tt poonam@rri.res.in}
} \\
{\normalsize\em Raman Research Institute, C. V. Raman Avenue,} \\
{\normalsize\em Bangalore 560 080, India. \vspace*{0.15cm}}
 }
\date{February 5, 2010}  %
\maketitle \thispagestyle{empty} \vspace{-0.8cm}
\begin{abstract}
\noindent In vacuum or constant density matter, the two flavor
neutrino oscillation formulae are insensitive to the presence of CP
violating phases owing to the fact that the CP phase can be gauged
away. In sharp contrast to the above case, we show that the CP
violating phases can not be gauged away in presence of adiabatically
changing background density accompanied by varying CP phases. We
present a pure geometric visualization of this fact by exploiting
Pancharatnam's prescription of cyclic quantum projections.
Consequently the topological phase obtained in Phys. Rev. D 79,
096013 (2009)  can become geometric if CP violation occurs in a
varying density medium.
\end{abstract}

\vspace{1.cm} \centerline{Pacs numbers: 03.65.Vf,14.60.Pq}
\vspace{.3cm} \centerline{Keywords: geometric phases, neutrino
oscillations, CP violation}

\newpage

\section{Introduction}
\label{intro} Pontecorvo's insightful idea of neutrino
oscillations~\cite{pontecorvo} has met phenomenal success over the
past several decades in explaining the observed deficit of neutrinos
from a wide variety of astrophysical
 and terrestrial sources. It is now well established that the different
neutrino flavors oscillate among themselves while conserving the
lepton number. The coherent
 phase picked up by the evolving states that shows up in transition
amplitudes leads to the phenomena of oscillations. The fact that
there is a geometric interpretation of the standard neutrino
oscillation formalism was recently shown~\cite{pmcpc} for the
specific case of CP conservation (CPC) and two flavors.

The topological\footnote{By our definition, the topological phase
refers to phase factors that are insensitive to small changes in the
circuit (and are invariant under deformations of circuit), while
geometric phases are sensitive to such changes. We use this
distinction between the two terms throughout this article (unless
otherwise specified).} phase obtained in Ref.~\cite{pmcpc} was
robust and did not depend upon the nature of the evolution.   We
pointed out that as long as the singular point in ray space was
encircled, the phase of $\pi$ remained irrespective of the
evolution being in vacuum or in medium. The $\pi$ anholonomy was
first noticed in Refs.~\cite{lh} and \cite{hlh} as a phenomenon
associated with real and symmetric Hamiltonians.
 The presence of this topological phase was ascribed to the structure of
 leptonic mixing matrix. This implied that there was no unexpected surprise
due to this topological phase and that the standard quantum
mechanical treatment takes into account Pancharatnam's phases
correctly (in fact is a realization of the same).

 The purpose of the present work is to extend our analysis~\cite{pmcpc}  to the
 CP violating (CPV) case while restricting ourselves to two neutrino
 flavors. We will explicitly find the conditions under which it is possible to destroy
 the robustness of the topological phase by the inclusion of CPV terms.
  We consider  the general CPV form of neutrino Hamiltonian for the
 case of two neutrino flavors and derive an expression for the oscillation
probability for an arbitrary medium with in the adiabatic
approximation. Then we show that the cross terms in the probability
have a purely geometric interpretation by employing Pancharatnam's
prescription of cyclic quantum collapses. By doing so we establish a
crucial geometric difference between the CPC  and the CPV case.
Having established our result, we briefly discuss our result in two
possible physical scenarios where  CPV can appear in the Hamiltonian
and leave its imprint at the level of detection probability.

The information on neutrino masses and mixings gleaned from most
oscillation experiments can be conveniently analysed considering two
neutrino situation or quasi-two-neutrino-situation (for example, the
one mass scale dominant approximation)~\cite{Akhmedov:2004ve}. So,
we focus only on two flavor case in the present work which
 provides a clean visualization of physical effects via rotations and
reflections on the \poi sphere.

This paper is organized as follows. In Sec.~\ref{sec2}, we describe
the general Hamiltonian for two flavor neutrinos, including the CPV
term. We also discuss possible sources of such a CPV term.
 Then in Sec.~\ref{sec3}, we obtain the neutrino oscillation formulae in
presence of  CPV term and establish  connection of the cross terms
appearing in the probability with the Pancharatnam's geometric
phase. We explicitly demonstrate (pictorially) the geometric
differences
 between the case of constant density  and adiabatically
changing matter background, both for CPC and CPV situations which is
the central result of this article. In this context, we also comment
on the action of CP and its
 connection with Wigner's theorem in Sec.~\ref{sec4}. We end with
 concluding remarks in
 Sec.~\ref{concl}.

\section{Hamiltonian for two flavor neutrinos}
 \label{sec2}
 It is well-known that the two flavor neutrino system is
equivalent to a two state quantum system~\cite{raffeltbook,pmcpc} in
the ultra-relativistic limit (for equal fixed momenta of the two
neutrinos). Upon linearizing the Dirac equation in the
ultra-relativistic limit, the propagation of neutrinos through
vacuum or matter can be described by an effective \schd-like
equation.

Let us examine the general hermitian\footnote{neglecting absorption}
form of  Hamiltonian for any two state
system~\cite{berrypancharatnam},
 \begin{eqnarray}
  \mathbb{H} &=& \hat r \cdot \vec\sigma + r_0 {\mathbb{I}}_2 \nonumber\\
 &=&
\begin{pmatrix}
 z  &  x- i y  \\
x+ i y & - z
 \end{pmatrix} + r_0 {\mathbb{I}}_2~, \nonumber\\
 &\simeq&  \begin{pmatrix}
-\cos \vartheta  &  \sin \vartheta~ e^{-i \varphi}  \\ \sin
\vartheta ~e^{i \varphi} &  \cos \vartheta
 \end{pmatrix}~,\label{eqham}
 \end{eqnarray}
where, $x$, $ y$  and $z$ are the three independent parameters
appearing in the Hamiltonian. The term proportional to Identity
$r_0$ leads to non-zero trace and adds an overall phase factor,
$e^{-i \int r_0 dt}$ to the evolving state. But the states have the
freedom of redefinition upto a phase so this extra phase does not
affect oscillations and we can safely omit this term and deal with
traceless Hamiltonian (last row of Eq.~(\ref{eqham}))  spanned by
the Pauli matrices. $\vartheta$ and $\varphi$ are the polar angles
and $\hat r$ is a unit vector. The eigenvalues of Eq.~(\ref{eqham})
are $ {\lambda}_\pm =  \pm \int_0^t \sqrt{x^2 + y^2 + z^2} ~dt' =
\pm 1$ and the normalized eigenstates are : \bea
\ket{\vartheta,\varphi,+} =\begin{pmatrix} \cos (\vartheta/2) e^{-i
\varphi}  \\ \sin (\vartheta/2)
 \end{pmatrix}~ &{\rm{and}}& ~
\ket{\vartheta,\varphi,-} = \begin{pmatrix} - \sin (\vartheta/2)
e^{-i \varphi} \\ \cos (\vartheta/2) \end{pmatrix}~, \eea
  The ray space in this case is a
two-dimensional sphere ${\mathbb {S}}^2$ which is known as \poi
sphere (also referred to as the  \bloch sphere in quantum
mechanics). Pictorially, oscillations are unitary rotations about an
axis on the \poi sphere while CP transformation can be represented
as reflection in the $x-z$ plane on the \poi sphere. The terms
appearing in the $\mathbb{H}$ due to vacuum, normal matter (with
standard interactions (SI) and non-standard interactions (NSI)) and
due to neutrino backgrounds with SI encountered by the neutrinos as
they propagate between the source and the detector are listed in
Table~\ref{table1}\footnote{The angle $\vartheta$ is connected to
the mixing angle $\Theta$ used in the standard neutrino oscillation
literature by $\vartheta=2 \Theta$.}. In general, oscillations in
vacuum or any medium can be viewed as the phenomenon of elliptic
birefringence in optics as explained in~\cite{pmcpc}.

%
%
\begin{table*}[t]{\scriptsize%
\begin{center}
\begin{tabular}{| l |  l | l | l | }
 \hline
&& & \\
\hspace{0.5cm}{\rm {Medium}}\hspace{0.5cm} & \hspace{0.5cm}{
{$x$}}\hspace{0.5cm} & \hspace{0.5cm}{ {$y$}}\hspace{0.5cm} &
\hspace{0.5cm}{ {$z$}}\hspace{0.5cm}
\\
&&&\\
\hline Vacuum          & $(\omega/2) \sin\vartheta$ & $0$ & -$(\omega/2) \cos\vartheta$\\
          \hline Normal &&&\\ matter+SI
          & $(\omega/2) \sin\vartheta$ & $0$ & -$(\omega/2) \cos\vartheta + V_C/2$
          \\
                    \hline Normal &&&\\ matter+NSI
          &$\re{ (\omega/2) \sin \vartheta +\ep_{ey}/2 }$  & $\im{ (\omega/2) \sin \vartheta +\ep_{ey}/2 }$
           & -$(\omega/2) \cos\vartheta + V_C/2 + (\ep_{ee}-\ep_{yy})/2 $\\
          \hline
          Neutrino &&&\\ backgrounds+SI
          & $\re{ (\omega/2) \sin \vartheta +B_{ey}/2 }$ &
          $\im{ (\omega/2) \sin \vartheta +B_{ey}/2 }$& -$(\omega/2) \cos\vartheta + V_C/2 + B/2 $ \\
          \hline
\end{tabular}
\caption[]{\footnotesize{The three independent elements of
$\mathbb{H}$ in different kinds of media - vacuum, normal matter
with SI, ordinary matter with NSI and mixed state neutrino
backgrounds (see text). } } \label{table1}
\end{center}
}
\end{table*}

Let us comment upon the CP properties of different media mentioned
in Table~\ref{table1} and appearance of CPV term in the Hamiltonian.
The action of CP can be understood in terms of $\mathbb{H} \to
\mathbb{H}^T$. Since only $\sigma_y$ changes sign under CP, CPV
implies that $y \neq 0$. Vacuum is CP symmetric which implies $y=0$.
The other two terms ($x$ and $z$) depend on mixing angle in vacuum
$\vartheta/2$ and the vacuum oscillation frequency $\omega = \delta
m^2 /2p$ where $\delta m^2$ is the mass-squared difference and $p
\simeq E$ is the neutrino momentum or energy in the
ultra-relativistic limit. Normal matter is CP asymmetric since
neutrinos and antineutrinos interact with matter in different ways
(Mikheyev-Smirnov-Wolfenstein~\cite{Mikheyev:1987qk,Wolfenstein:1977ue}
effect can enhance oscillation in the neutrino channel and suppress
in the antineutrino channel) but due to the absence of flavor
changing neutral currents in the Standard Model (SM) there are no
off diagonal terms and the Hamiltonian still preserves CP ($y=0$) in
the absence of intrinsic CPV phases. An important consequence of
this is that if vacuum mixing and neutrino masses are zero, then
neutrinos can not change flavor.
  In Table~\ref{table1}, the second row corresponds to this case.  $V_C=\sqrt{2} G_F n_e$ is
the charged current potential due to coherent forward scattering of
$\nu_e$ with electrons in matter. $G_F$ is the Fermi coupling
constant and $n_e$ is the electron number density.

The minimal theoretical scenario needed to describe neutrino flavor
oscillations is the requirement of neutrino masses, for which there
is concrete evidence from various oscillation experiments. It is
well-known that the SM by itself predicts massless neutrinos. The
simplest way is to add dimension five non-renormalizable terms
consistent with symmetries and particle content of the SM which
leads to desired Majorana masses for the left-handed neutrinos.
But, the neutrino interactions involving the light fields are still
assumed to be described by weak interactions within the SM and this
is  minimal SM extension for accomodating neutrino mass.
 However, once we invoke new physics in
order to explain the non-zero neutrino masses, it seems rather
unnatural to leave out the NSI which allow for flavor changing
interactions as well as are new sources of CP violation which can
affect production, detection
 and propagation of neutrinos~\cite{grossman-1995-359,iss}. Some of the early
 attempts discussing new sources of lepton flavor violation (for instance,
 R-parity violating supersymmetry) were geared
 towards providing an alternate explanation for the observed deficit of
 neutrinos coming from the Sun in the limiting case of zero neutrino mass
 and absence of vacuum mixing~\cite{valle,guzzo,roulet}.
In recent years, the emphasis has shifted towards understanding the
interplay between SI and NSI and whether future oscillation
experiments can test these NSI apart from determining the
oscillation parameters precisely. This has led to an upsurge in
research activity in this
direction~\cite{datta1,concha,mehtalq,2fnsi,wintercpvnsi,densensi,meloni,gago,biggio}.
 Non-standard physics which allows for flavor changing interactions
  may also lead to additional CPV phases which appear as $y \neq 0$ in
 Eq.~(\ref{eqham}).

The first possibility to have $y \neq 0$ is the case of  NSI in
propagation (which affects the coherent forward scattering of
neutrinos with background matter)\footnote{We focus only on NSI
during propagation of neutrinos through matter. However it should be
noted that NSI can also affect production and detection processes as
 discussed in Refs.~\cite{datta1,mehtalq,concha,biggio}.} which induces new CPV
 phases. Such extrinsic\footnote{The term extrinsic refers to matter-induced CPV
phases.} CPV phases can mimic the signature of intrinsic CPV phases
thereby affecting the determination of intrinsic phase appearing in
the leptonic mixing matrix. In the three flavor case, the problem
becomes very complicated and one needs to carefully disentangle the
extrinsic CPV from the intrinsic one by looking at the CPV
observables~\cite{wintercpvnsi}. Here in the two flavor case, there
are no intrinsic phases~\cite{yossitasi91} but extrinsic CPV phases
can actually appear at the level of Hamiltonian.
 If we consider only those operators that
arise at a scale much lower than lepton number violating scale then
the relevant interaction is \bea {\cal L} = \sum_{{f};
~{\alpha,\beta}} 4 \dfrac{G_F}{\sqrt 2} \bar \nu_{\alpha L} \gamma
^\mu \nu_{\beta L} \left( \epsilon_{\alpha \beta}^{fL} \bar f_L
\gamma_\mu f_L + \epsilon_{\alpha \beta}^{fR} \bar f_R \gamma_\mu
f_R \right)~,\eea where $f=e,p,n$ and $\alpha,\beta=e,\mu,\tau$.
Since the scale at which this interaction arises is supposed to be
not too far from the electroweak scale $\Lambda_{ew} \sim
{G_F}^{-1/2}$, its coupling may be parameterized by $ \epsilon_{\al
\be} G_F$, where $\epsilon \sim ({\Lambda_{ew}}/{\Lambda_{NP}})^2$.
The terms appearing in the flavor Hamiltonian in presence of
ordinary matter and NSI are given in third row of
Table~\ref{table1}.
At the level of underlying Lagrangian describing NSI, the NSI
coupling of the neutrino can be to $e,u,d$ or $e,p,n$. But from
phenomenological point of view, only the sum (incoherent) of all
these individual contributions is relevant. The  effect of coherent
forward scattering induced by such interaction terms in Lagrangian
on the propagation of neutrinos in ordinary neutral unpolarized
medium is governed by the parameters, \bea \epsilon_{\al \be} =
\sum_{f=e,u,d} \dfrac{n_f}{n_e} \epsilon^f_{\al \be}~,
\eea where $n_f$ is the density of fermion $f$ in medium crossed by
the neutrino. Also, $\epsilon^f= \epsilon^{fL} + \epsilon^{fR}$.

Note that a CPV term in the Hamiltonian can have its origin in
different sources,
 for example by inclusion of non-standard neutrino-matter
 interactions during propagation as described above
 or via CPV neutrino background which is encountered by neutrinos
 emanating from
 a dense supernova core. In the latter case, if the background neutrinos are in
 a mixed~\footnote{Not an impure state in the
 conventional sense.} state given by $\ket{\nu_b}=\gamma_e \ket{\nu_e} + \gamma_y \ket{\nu_y}$,
 CPV can arise in the two flavor neutrino Hamiltonian
 without the need to invoke any new
 physics~\cite{smirnovtalk,dighetalk1}. This is due to neutrino self
 interactions~\cite{thomson,Pantaleone:1992eq,qianfuller}.
The pure state flavor density matrix for a given momentum mode $\bp$
is \beq \varrho_\bp = \ket{\nu_b}\bra{\nu_b}= \begin{pmatrix}
   |\gamma_e|^2  &  \gamma_e \gamma_y ^\star \\
\gamma_e^\star \gamma_y& |\gamma_y|^2 \end{pmatrix}~. \eeq
  Background neutrinos will be in mixed state naturally if they also
   undergo oscillations.
The terms of fourth row in Table~\ref{table1} are given by
  \bea B &=& \sqrt{2} G_F \int d^3 {\bq} (1-\cos \theta_{\bp\,\bq})
[(\varrho_\bq-\bar\varrho_\bq)_{ee} - (\varrho_\bq-\bar\varrho_\bq)_{yy} ]~, \nonumber\\
B_{ey} &=& \sqrt{2} G_F \int d^3 { \bq} (1-\cos \theta_{\bp\,\bq})
[(\varrho_\bq-\bar\varrho_\bq)_{ey}  ]~, \nonumber\\ B_{ye} &=&
\sqrt{2} G_F \int d^3 { \bq} (1-\cos \theta_{\bp\,\bq})
[(\varrho_\bq-\bar\varrho_\bq)_{ye} ]~, \eea where
$\theta_{\bp\,\bq}$ is the angle between the direction of
propagating neutrino with momentum $\bp$ and the direction of other
neutrinos in the ensemble with momentum label $\bq$. The quantity
${\varrho_{\bq{ey}}} ({\bar \varrho_{\bq {ey}}})$ stands for the
matrix element of the density matrix operator,
$\me{\varrho_{\bq}}{\nu_e}{\nu_y}
(\me{\varrho_{\bq}}{\bar\nu_e}{\bar\nu_y})$. Hence for mixed state
background neutrinos, all the three terms of the Hamiltonian $x$,
$y$ and $z$ can be non-zero in contrast to the case of vacuum or
ordinary matter with SI. Thus we note that the Hamiltonian with CPV
neutrino background requires all the three parameters for its
complete description. The background potential matrix is no longer
diagonal and therefore even massless neutrinos can oscillate.

We should emphasize that within the SM, the  only situation where
one can get flavor off-diagonal terms in the Hamiltonian without any
new physics is the case of dense neutrino backgrounds. Here
off-diagonal terms arise not due to any new flavor changing
interactions but due to background neutrinos being in a mixed flavor
state. If the background neutrinos were in pure flavor state then
the off-diagonal terms in the potential matrix would disappear and
one ends up with a CPC Hamiltonian as in the case of ordinary matter
when only SI are taken into account. Hence such a mixed state
neutrino background provides us with a minimal scenario for CPV
effects to play a role. In addition, invoking NSI in the case of
dense neutrino backgrounds leads to extra source of CPV phases in
the Hamiltonian~\cite{densensi} over and above those induced by
mixed state neutrino background.

The question then arises  if there is any visible consequence of
extrinsic CPV phase or $y \neq 0$ at the level of probability in the
two flavor case whatever its source might be. It is well-known that
in the case of CPV medium with constant density, this term is of no
consequence. Below we will first establish this fact by showing that
the extra phases in the mixing matrix can be gauged
away~\cite{yossitasi91}. By
  using Pancharatnam's prescription of cyclic quantum collapses we show that the geometric phase
  is always restricted to be topological (two pairs of orthogonal states can  always be made to lie on a
  great circle)~\cite{pmcpc}. Further we will  discuss the case of varying density CPV
 medium where  it is possible to have an observable effect of CPV phases at the probability level.
 This  is connected intimately to the fact that the Pancharatnam's phase can become
 geometric (different from $\pi,0$) in such a situation. We give a pure geometric view of this effect by
  using Pancharatnam's prescription~\cite{Pancharatnam:1956}.

\section{CPV phases and the two flavor oscillation probability}
\label{sec3} Solving the full neutrino evolution equation in matter
exactly is
 formidable analytically  because of the fact that density in general is not
constant but varies in space (or time in natural units). Usually one
assumes adiabaticity : the variation of density is small over one
oscillation length. This is the quasi-static
approximation~\cite{yossi}.

If we start with the most general form of mixing matrix which
includes the CPV phases, we can argue that only one angle is
sufficient to parameterize this matrix and the extra phase can
always be gauged away by appropriate redefinition of the flavor
states~\cite{yossitasi91,fukugita,kim}. This can be shown explicitly
as follows \bea \label{umat} {\mathbb{U}} &=& {\mathbb{U}}_{\nu_L}
{\mathbb{U}}_{l_L}^\dagger = \begin{pmatrix} \cos(\vartheta/2)
e^{i \alpha} &  \sin (\vartheta/2) e^{i \beta} \\
 -\sin (\vartheta/2) e^{i \gamma} & \cos (\vartheta/2) e^{i (-\alpha+\beta+\gamma)}
 \end{pmatrix}~,\nonumber\\ \eea
 If we define
\bea {\mathbb{P}}_\nu = \begin{pmatrix}
 e^{-i \alpha} &  \\
 & e^{-i \gamma}
 \end{pmatrix}~, \quad && {\mathbb{P}}_l =
\begin{pmatrix}
 1 &  \\
 & e^{i (-\alpha+\beta)}
 \end{pmatrix}~,\label{gauge} \eea
then under a transformation ${\mathbb{U}} \to {\mathbb{P}}_\nu
{\mathbb{U}} {\mathbb{P}}_l^\star $, we can eliminate the three
phases from the mixing matrix. The mass eigenstates are defined up
to a phase and  therefore we can redefine them $\nu _{L,R} \to
{\mathbb{P}}_\nu \nu_{L,R}$ (Note that $\nu_{L}$ is denoted by
$\ket{\vartheta_i,\varphi_i,\pm}$ where $i=1,2$ runs over the number
of generations) and similarly for $l_{L,R} \to {\mathbb{P}}_l
l_{L,R}$. With the above transformation, the mass matrices remain
real and unchanged. This leads to no observable CPV (Dirac type) in
the two flavor case. But, a Majorana type CPV phase can survive even
in the two flavor case~\cite{doi,fukugita,kim,giunti}. Note that in
certain extended models, such as left right symmetric models and
supersymmetric models, it is possible to have CPV only with two
flavors. The CPV characteristic of Majorana phases can be probed in
effects connected with the neutrino mass term such as neutrinoless
double beta decay~\cite{doi}. But the main point is that neutrino
oscillation experiments are insensitive to Dirac or Majorana nature
of neutrinos~\cite{fukugita,giunti}.
 So, it suffices to take the
mixing matrix to be real, \bea \label{umatreal} {\mathbb{U}} &=&
\begin{pmatrix}
\cos(\vartheta/2) &  \sin (\vartheta/2) \\
 -\sin (\vartheta/2)  & \cos (\vartheta/2)
\end{pmatrix}~, \eea without any loss of generality. This leads to no
observable significance of intrinsic CPV phases as far as two flavor
neutrino oscillation formalism is concerned~\cite{giunti}. Note that
at any instant, we can always get rid of the intrinsic CPV phases in
the mixing matrix.

However in case of an adiabatic evolution, is it still possible to
define a gauge transformation which can gauge away the dependence of
CPV phases. Let us ask if the transition probability for the two
flavor neutrinos is at all sensitive to the CPV phases in presence
of adiabatic evolution. In what follows, we consider neutrinos
propagating through an arbitrary medium and obtain the expression
for transition probability and analyse the effect of CPV phase in
the Hamiltonian in Eq.~(\ref{eqham}).

To keep the discussion general, we start with a neutrino created as
a  flavor state $\ket{\nu_\alpha}$ and detected as  $\ket{\nu_\be}$.
  The state $\ket{\nu_\al}$ is
\bea \ket{\nu_\al} = \nu_{\al,+} \ket{\vartheta_1,\varphi_1,{+}} +
\nu_{\al,-} \ket{\vartheta_1,\varphi_1,{-}}~,\label{psi}\eea where
$\ket{\vartheta_1,\varphi_1,{\pm}}$ denote orthogonal pair of energy
eigenstates of the Hamiltonian $\mathbb{H} (\vartheta_1,\varphi_1)$
(Eq.~(\ref{eqham})). The two orthogonal mass states
$\ket{\vartheta_1,\varphi_1,{\pm}}$ evolve into
$\ket{\vartheta_2,\varphi_2,{\pm}}$ as a result of adiabatic
evolution~\cite{yossi}
\begin{eqnarray}
 \ket{\vartheta_1,\varphi_1,{\pm}} &\to&
 e^{-
i {\mathcal D}_\pm} \ket{\vartheta_2,\varphi_2,{\pm}}~,
\label{correctadiab}
\end{eqnarray}
with
$ {\mathcal D}_\pm =  \pm   \int_0^t \sqrt{x^2 + y^2 + z^2} ~dt' +
\int_{0}^t r_0~ dt' $  as the dynamical phases obtained from
Eq.~(\ref{eqham}) where $x,y,z$ for specific medium can be read off
from Table~\ref{table1}.

This is the zeroth order adiabatic approximation. However the two
states can also pick up a geometric contribution as they traverse
the region between the source and the detector~\cite{pmcpc}. In
general the geometric component picked up by the individual
eigenstates under adiabatic evolution vanishes on account of
evolution being along the geodesic curve\footnote{Except when the
evolution along the geodesic curve happens to cross the antipodal
point and the geometric phase then becomes $\pi$.}. But, a net
geometric phase can appear if we can perform interference experiment
between these energy eigenstates in the energy space. Neutrinos are
produced and detected as flavor states which are superpositions of
mass states, so this condition of realising interference in energy
space is automatically achieved~\cite{pmcpc}. The appearance of
geometric phase can be understood from the description of vertical
lift~\cite{sam} which accounts for the net non-zero geometric phase
during any evolution (closed or open). Note that open loops can also
be closed by connecting them by the shorter
 geodesic. The geometric phase is determined by the path of
 the state vector on the \poi sphere.

\begin{figure*}
\hskip.0in
\includegraphics[width=6cm,height=6cm]{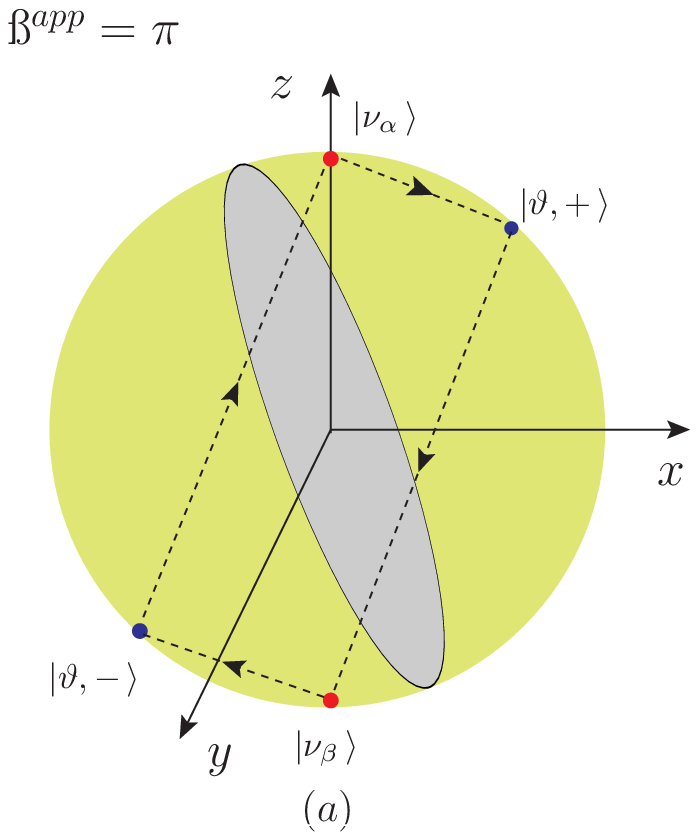}
 \vskip -6cm \hskip 3in
\includegraphics[width=6cm,height=6cm]{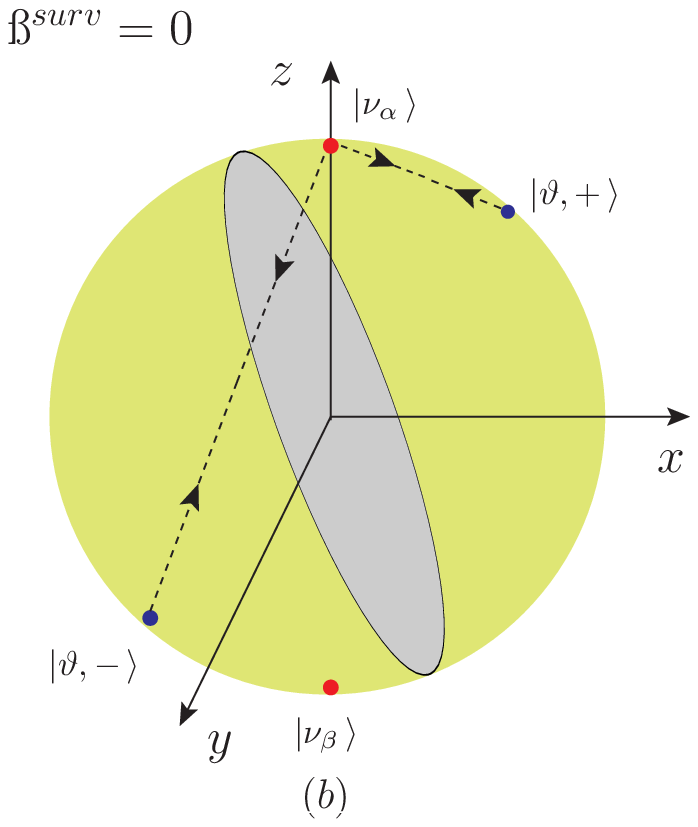}\\
\hskip-3in
\includegraphics[width=6cm,height=6cm]{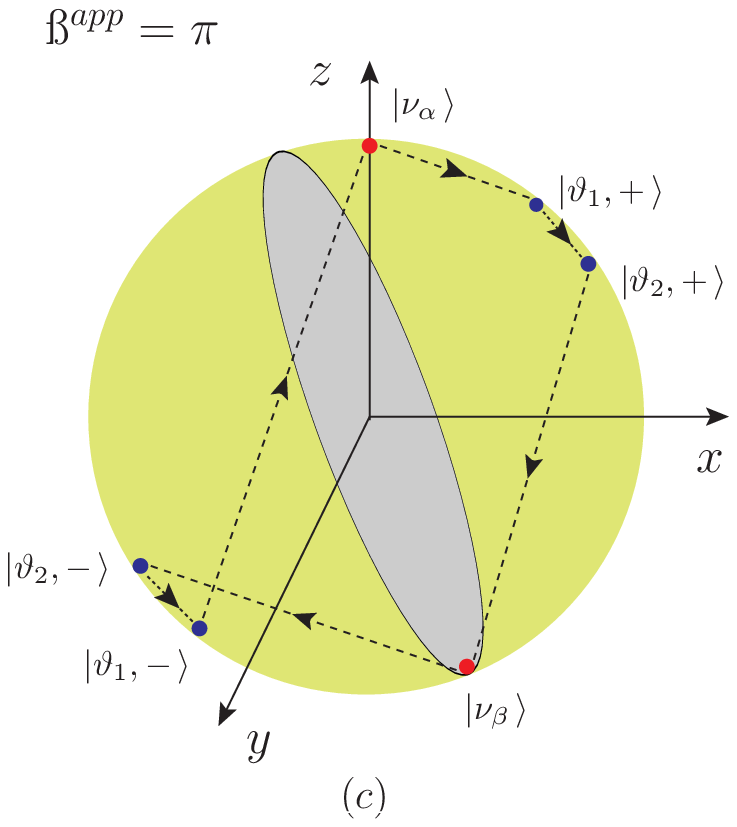}
 \vskip -6cm \hskip 3in
\includegraphics[width=6cm,height=6cm]{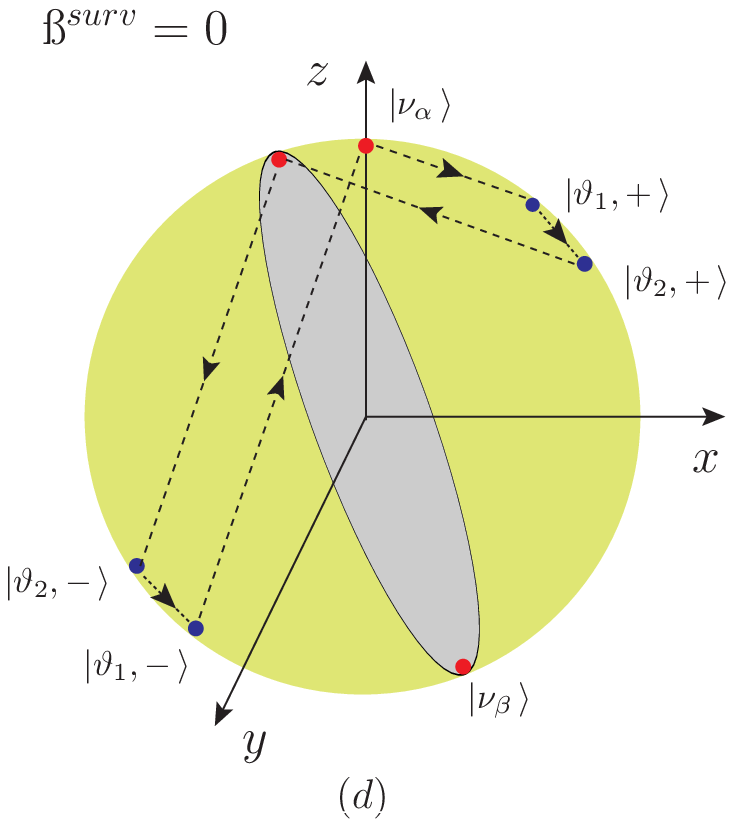}
\caption{\footnotesize{Pictorial depiction of the cross terms
(Eq.~\ref{prob}) in case of appearance and survival probability for
the CPC case. The direction of collapse processes connecting the
four (six) neutrino states  appearing in constant (varying) density
case in the cross term  are represented on the \poi sphere by
straight lines with arrows (see text). Note that the CPC mass states
are labelled by  $\ket{\vartheta_i,{\pm}}$ ($i=1,2$).
 Cases (a) and (b) correspond to constant density  situation  while (c) and (d)
 stand for varying density situation.
 The states $\ket{\vartheta_2,{\pm}} $ are
obtained after adiabatically evolving $\ket{\vartheta_1,{\pm}}$.
Adiabatic evolution between the mass states is shown by a dotted
line while dashed lines refer to collapse processes.
 }}
\label{fig0}
\end{figure*}
\begin{figure*}
\hskip.0in
\includegraphics[width=6cm,height=6cm]{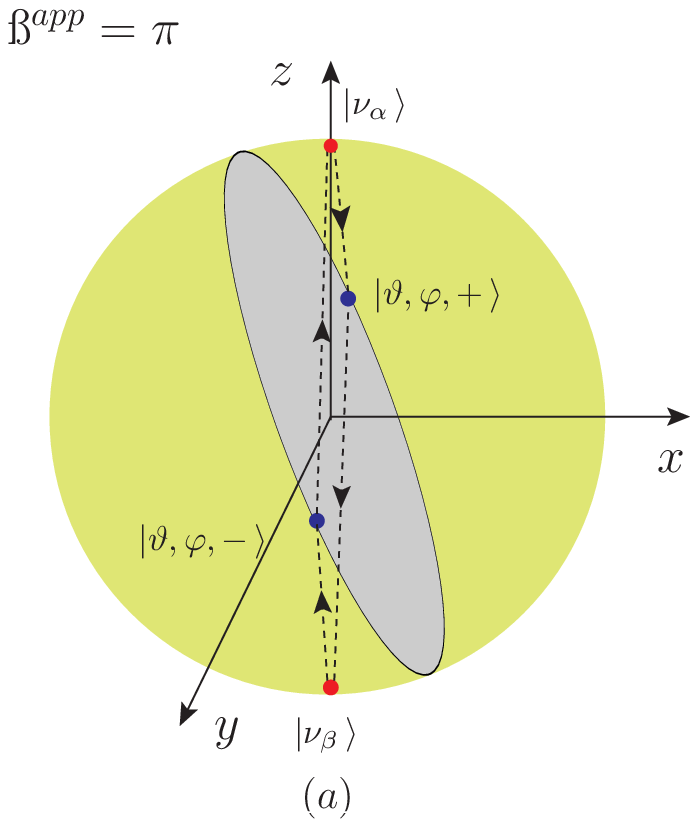}
 \vskip -6cm \hskip 3in
\includegraphics[width=6cm,height=6cm]{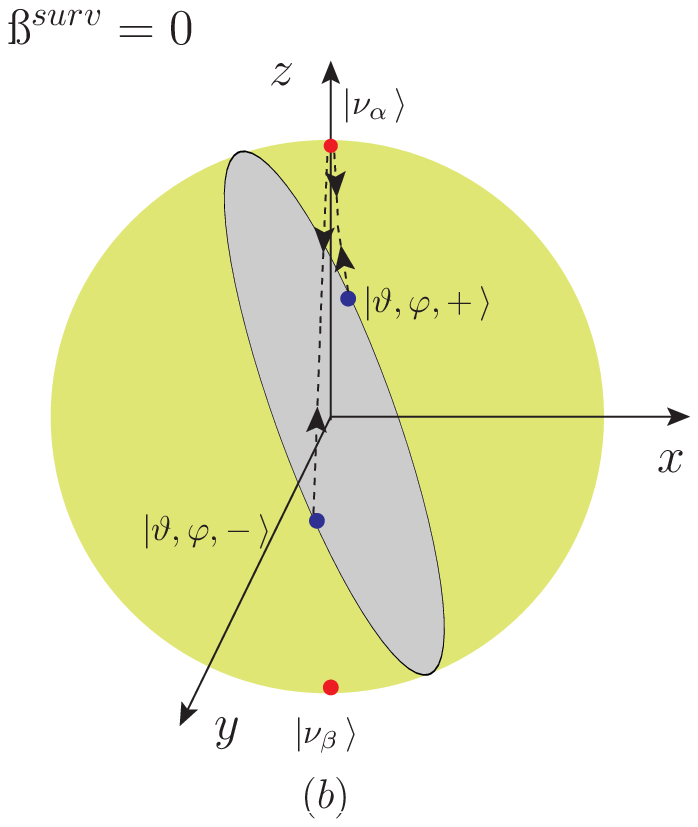}\\
\hskip-3in
\includegraphics[width=6cm,height=6cm]{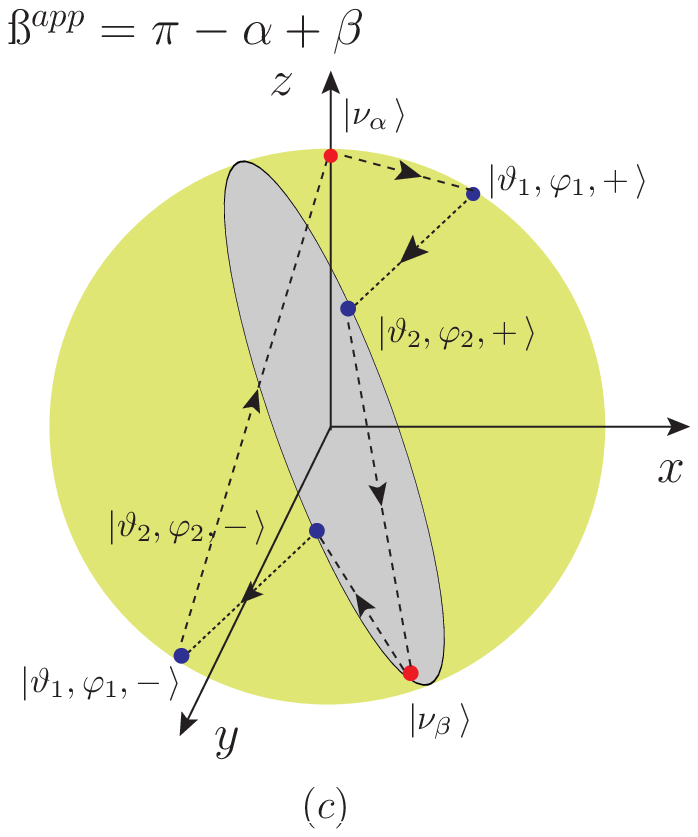}
 \vskip -6cm \hskip 3in
\includegraphics[width=6cm,height=6cm]{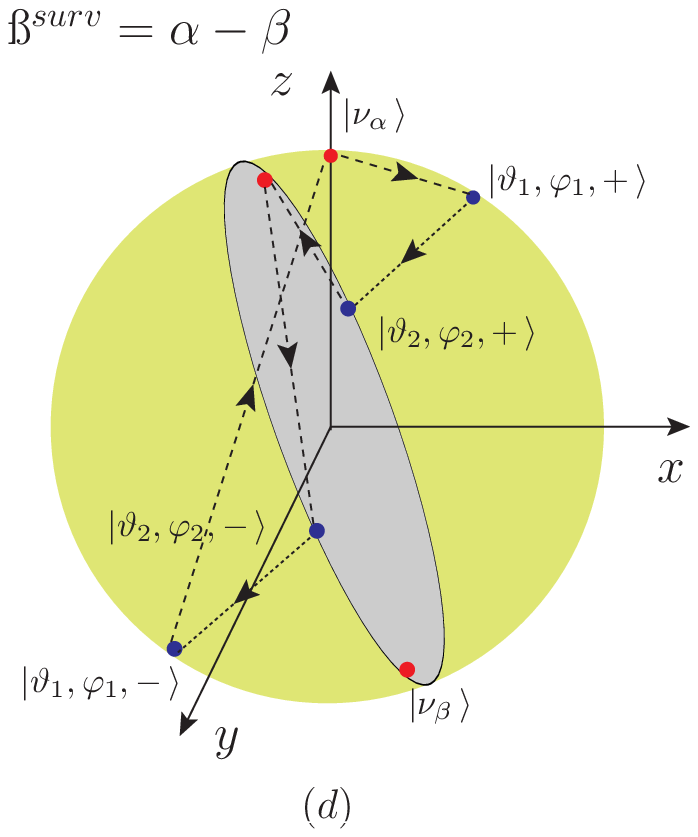}
\caption{\footnotesize{Direction of collapse processes connecting
the four (six) neutrino states in absence  (presence) of variation
of density (and CPV phases) corresponding to the cross terms in
probability (Eq.~(\ref{prob})) are represented on the \poi sphere by
straight lines with arrows (see text). Note that the in presence of
CPV,
 $\ket{\vartheta_i,\varphi_i,{\pm}} (i=1,2)$ are used to denote the mass states.
 Cases (a) and (b) correspond
to appearance and survival probability in a CPV ($\varphi \neq 0$)
but constant density  situation.
 Cases (c) and (d) stand for appearance and survival probability in
 presence of CPV and varying density case. Adiabatic evolution between the
mass states is shown by a dotted line while dashed lines refer to
collapse processes.
 }}
\label{fig1}
\end{figure*}

 The states $\ket{\vartheta_1,\varphi_1,{\pm}}$ and
$\ket{\vartheta_2,\varphi_2,{\pm}}$ are
 connected by parallel transport rule
on the \poi sphere. The two time evolved states $e^{- i {\mathcal
D}_\pm} \ket{\vartheta_2,\varphi_2,{\pm}}$ form the final flavor
state $\ket{\nu_\be}$ at the detector \bea
 \ket{\nu_\be} = \nu_{\be,+} \ket{\vartheta_2,\varphi_2,{+}} +
\nu_{\be,-} \ket{\vartheta_2,\varphi_2,{-}}~.\eea Note that the
initial and final flavor states  are shown as red bullets in
Figs.~\ref{fig0} and \ref{fig1}. $\ket{\vartheta_i,\varphi_i,{\pm}}
(i=1,2)$ correspond to the orthogonal pairs of mass (energy)
eigenstates lying on distinct elliptic axes depending upon the
values of $\vartheta_i,\varphi_i$ (shown as blue bullets in
Figs.~\ref{fig0} and \ref{fig1}).

To illustrate this point of view, let us carefully examine the
expression for transition probability along the lines of
Ref.~\cite{pmcpc}. The probability for flavor transition ${\nu_\al}
\to {\nu_\be}$ is given by squaring the amplitude,
\bea {\cal P} (\nu_\al \to \nu_\be) &=& |{\cal A} (\nu_\al \to
\nu_\be) |^2 \nonumber\\ &=&
    \scx{\nu_\al}{\vartheta_1,\varphi_1,{+}}
\scx{\vartheta _2,\varphi_2,{+}}{\nu_\be}
\scx{\nu_\be}{\vartheta_2,\varphi_2,{+}}
\scx{\vartheta_1,\varphi_1,{+}}{\nu_\al}
\nonumber\\ &+&
 \scx{\nu_\al}{\vartheta_1,\varphi_1,{-}} \scx{\vartheta
_2,\varphi_2,{-}}{\nu_\be}
\scx{\nu_\be}{\vartheta_2,\varphi_2,{-}} \scx{\vartheta_1,\varphi_1,{-}}{\nu_\al} \nonumber \\
 &+&
   [\scx{\nu_\al}{\vartheta_1,\varphi_1,{-}}e^{ i {{\cal D}_-}}  \scx{
 \vartheta
_2,\varphi_2,{-}}{\nu_\be} \scx{\nu_\be}{\vartheta_2,\varphi_2,{+}}
e^{ -i {\cal D}_+} \scx{\vartheta_1,\varphi_1,{+}}{\nu_\al} +
{\rm{c.c.}} ]~. \label{prob} \nonumber\\ \eea
The  two direct terms  correspond to classical probability and two
cross terms contain the effect of geometric and dynamical phases.
 In the limit when coherence is lost (for instance when the path
length is much larger than the oscillation length), the total
probability reduces to the classical probability. The quantum
mechanical effect of oscillation essentially lies in the cross
terms.

At this juncture, we would like to introduce Pancharatnam's idea of
closed loop quantum collapses and the related geometric
interpretation of the cross terms appearing in the expression of
probability (Eq.~(\ref{prob})).
 Given any three rays $\ket{\mathfrak{A}}$, $\ket{\mathfrak {B}}$ and $\ket{\mathfrak {C}}$
in the projective Hilbert space for a two state system (such that
the neighbouring rays are non-orthogonal) the phase of the complex
number $\scx{\mathfrak A}{\mathfrak C} \scx{\mathfrak C}{\mathfrak
B} \scx{\mathfrak B}{\mathfrak A} $ is given by $\Omega/2$ where
$\Omega$ is the solid angle subtended by the geodesic triangle at
the center of the \poi sphere. This excess phase is known as
{\it{the Pancharatnam phase}}. Pancharatnam's phase reflects the
curvature of ray space
 and is independent of any parameterization or slow variation.
Thus it can also appear in situations where the Hamiltonian is
constant in time i.e. for the case of vacuum or constant density
matter in the case of neutrino oscillations. Furthermore, note the
fact that \sch evolution (possibly) interrupted by measurements can
lead to Pancharatnam's phase. The crucial requirement for
Pancharatnam's phase to be well-defined is to have cyclic projection
of (at least three) states and the consecutive collapses should be
between  nonorthogonal rays.

 In  Eq.~(\ref{prob}) the cross term is related to
  the two path interferometer in energy space shown in Ref.~\cite{pmcpc}.
The structure of the cross term (without the dynamical evolution) is
a series of cyclic quantum collapses with intermediate adiabatic
evolutions given by $\ket{\nu_\al} \to
\ket{\vartheta_1,\varphi_1,{+}} \to \ket{\vartheta_2,\varphi_2,{+}}
\to \ket{\nu_\be} \to \ket{\vartheta_2,\varphi_2,{-}} \to
\ket{\vartheta_1,\varphi_1,{-}} \to \ket{\nu_\al} $ which
essentially maps a trajectory (as illustrated in Figs.~\ref{fig0}
and \ref{fig1}) and subtends a solid angle of $\Omega$ at the center
of the \poi sphere. The center of the \poi sphere is a singular
point which is the same as the degeneracy point corresponding to the
null Hamiltonian. So, the appearance of geometric phase is
intimately connected to the existence of this singular point.
%

The interference term (without the dynamical phase) can be expressed
as $r e^{i\ss}$ and therefore picks up a phase which will be
$\ss=\Omega/2$ (half the solid angle) due to Pancharatnam's
prescription. Note that even though the collapse process leads to a
loss in probability ($r \neq 1$) the phase of the cross term remains
unaltered whether we perform continuous evolution (unitary) or
collapses (non-unitary) as long as it is cyclic.

The Pancharatnam geometric phase is given by \bea \ss & = &
{\rm{Arg}} [\scx{\nu_\al}{\vartheta_1,\varphi_1,{-}} \scx{
 \vartheta
_2,\varphi_2,{-}}{\nu_\be} \scx{\nu_\be}{\vartheta_2,\varphi_2,{+}}
\scx{\vartheta_1,\varphi_1,{+}}{\nu_\al} ] \eea

A few remarks concerning the states $\ket{\nu_\al}$ and
$\ket{\nu_\be}$ are in order. The states  $\ket{\nu_\al}$ and
$\ket{\nu_\be}$ can be any general flavor states. Pure flavor states
($\ket{\nu_e}$ and $\ket{\nu_\mu}$) are very special (real) linear
superposition of mass eigenstates and correspond to those produced
and detected via weak interactions. If either $\ket{\nu_\al}$ or
$\ket{\nu_\be}$ are identical to any one of the four eigenstates
involved before and after the adiabatic evolution, the interference
term vanishes and then it is meaningless to talk about the phase of
such a term, whether dynamical or geometric. It is easy to
understand this from Pancharatnam's definition of parallelism
between any two nonorthogonal states.
%

%

For the case of {\bf{appearance  probability}} ${\cal P}$
(${\nu_\al} \to {\nu_\be}$) one gets a geometric phase  $\ss^{app}=
\Omega/2$.
 For $\nu_\be = \nu_\al$, i.e. {\bf{survival probability}} ${\cal P}$
(${\nu_\al} \to {\nu_\al}$) of the same flavor,
 it is easy to see that the path formed by cyclic collapses will again
 form a closed trajectory and subtend a non-zero solid angle at the
 origin.
 Therefore the interference term in survival probability will pick up a geometric phase which
 can be computed once we know the geometric phase in the
 case of appearance  probability  to
 be $\Omega/2$ by imposing unitarity
 which means that the interference terms should have a relative
sign. Thus  the cross term in survival probability will pick up a
phase given by
  $e^{i \ss} = e^{i(\pi-\Omega/2)}$. Hence both appearance and
  survival probabilities will pick up different geometric phases
 \beq
 \label{geo}
 \ss^{app}  = \Omega/2 \quad {\rm{and}} \quad \ss^{surv} = \pi-\Omega/2~.
 \eeq
Next we discuss the distinct physical situations that can arise. In
Figs.~\ref{fig0} and \ref{fig1}, we  pictorially depict the
geometric  differences in cross terms between the CPC and CPV cases
both for constant  and varying density situations. The CPC case was
discussed in Ref.~\cite{pmcpc}. In the present study with CPV
phases, the density and CPV phase are two parameters which can
either be constant or varying with distance, we list three possible
cases :

\begin{itemize}
\item[] \underline{\it{(A). Constant density case :-}}  The pictorial depiction of the
  direction of collapse processes is shown in Fig.~\ref{fig1}. In this case,
  we have four states (the two pair of mass and flavor states being
orthogonal) that will lie on a great circle as shown in
Fig.~\ref{fig1} (a) and (b). By appropriate
 gauge transformations the great circle can be made to coincide with
 the equatorial
 great circle.  The bottomline is that we can use real form of the
 leptonic mixing matrix (Eq.~(\ref{umatreal})) and
 the phase will remain topological as in the CPC case~\cite{pmcpc}.
The Pancharatnam geometric phase is given by \bea \ss &=&
 {\rm{Arg}}[\pm\cos^2 (\vartheta/2) \sin^2 (\vartheta/2)] = \pi,\,0~.
 \eea
 Using Pancharatnam's prescription of quantum collapses, we have shown that
in the two flavor context, {\it{CPV phase leaves no imprint of its
existence in oscillation probability for the case of constant
density matter}}, hence providing a {{purely geometric view of this
result}}.

It is then natural to ask if CPV plays any relevant role in presence
of adiabatic evolution  or alternatively one can ask if it is
possible to gauge away the CPV phase which itself is evolving in
time (or distance). We again use Pancharatnam's prescription to
analyze this situation.

\item[] \underline{\it{(B). Density and CPV phase varying with distance
:-}} In this case, we have the complex Hamiltonian
(Eq.~(\ref{eqham})) and the eigenstates will not be real in general
so the ray space is the full \poi sphere ${\mathbb S}^2$ (instead of
a great circle ${\mathbb S}^1$) in contrast to the case (A) listed
above (see Fig.~\ref{fig1} (c) and (d)).
There are four inner products appearing in the cross term,
$\scx{\nu_\al}{\vartheta_1,\varphi_1,{-}}$, $  \scx{
 \vartheta
_2,\varphi_2,{-}}{\nu_\be} $, $
\scx{\nu_\be}{\vartheta_2,\varphi_2,{+}}$
 and $\scx{\vartheta_1,\varphi_1,{+}}{\nu_\al}$ which are connected to the
 elements of the mixing matrix~\cite{pmcpc}. We saw that by a suitable gauge transformation,
 the mixing matrix can be made real for two flavor case (Eq.~(\ref{umatreal})).
 This implies that $\ket{\nu_\al}$ and $\ket{\vartheta_1,\varphi_1,\pm}$ can be chosen to be real and lying
 on the equatorial great circle ($x$-$z$ plane).
 However once this transformation is carried out, we do not have
 the freedom to choose the phases of the mass and flavor states any further. In general,
 the adiabatically evolved mass states  $\ket{\vartheta_2,\varphi_2,\pm}$ will be complex.
 Also,  the final flavor state  $\ket{\nu_\be}$ which will be a linear superposition
  of states $\ket{\vartheta_2,\varphi_2,\pm}$ with complex coefficients given by  Eq.~(\ref{umat}).
  The three states
 $\ket{\nu_\be}$ and $\ket{\vartheta_2,
 \varphi_2,\pm}$ will also lie on a different great circle and this
 is  clearly illustrated in Fig.~\ref{fig1} (c) and (d). The fact that the two great circles (one containing
 the initial flavor and mass states and the other one containing the final flavor and mass states) do
 not coincide will lead to a net geometric phase (which differs from topological phase obtained for the
  CPC case~\cite{pmcpc}).

Let us compute explicitly the cross term for this case. Two of the
inner products are real and positive, i.e.,
$\scx{\nu_\al}{\vartheta_1,\varphi_1,{-}} = \sin \vartheta_1$ and
$\scx{\vartheta_1,\varphi_1,{+}}{\nu_\al} = \cos \vartheta_1$. But
the other two are
 complex, given by
 $  \scx{
 \vartheta
_2,\varphi_2,{-}}{\nu_\be} = \cos \vartheta_2 e^{i
(-\alpha+\beta+\gamma)}$ and
 $ \scx{\nu_\be}{\vartheta_2,\varphi_2,{+}} = -\sin \vartheta_2 e^{-i\gamma}$.
 This implies that in general the geometric phase picked up will be given by
 \bea
 \ss^{app} = \pi - (\alpha - \beta) \quad &{\rm{and}}& \quad \ss^{surv} = \pi - \ss ^{app} = (\al -\be)~.
 \eea
 In the limit, $(\al-\be) \to 0$, the phase becomes topological and
 that would actually correspond to a constant density (CPC or CPV) situation.

 Geometrically,
 we see that during the course of adiabatic evolution the mass
 state $\ket{\vartheta_2,\varphi_2,\pm}$
gets lifted from one plane containing
$\ket{\vartheta_1,\varphi_1,\pm}$ and  thus explore the full ray
space instead of a great circle (with
$\scx{\vartheta_2,\varphi_2,\pm}{\vartheta_1,\varphi_1,\pm}$ being
complex in general). We have shown above that when presence of CPV
induced by background is coupled with an adiabatic evolution with
varying background density, the cross term of the probability (see
Eq.~(\ref{prob})) will depend upon the extent of CPV or variation of
$\varphi_i$ under adiabatic evolution ($\varphi_2 \neq \varphi_1$).
This leads to a geometric phase. Turning this argument around, we
can conclude that {\it{CPV in the Hamiltonian can  cause a deviation
of $e^{i \ss}$ from $\pm 1$  in presence of an adiabatic
evolution}}.

\item[]
\underline{\it{(C). Density varying and CPV phase constant :-}} In
this case, all  the six states involved in the collapse process with
intermediate adiabatic evolutions will be confined to a great circle
 and the phase will be topological in nature. This will be identical to the  CPC
 situation.

 \end{itemize}

The key result of this article can be stated as follows. The
topological phase obtained in Ref.~\cite{pmcpc} for CPC situation
was robust and did not depend up on any details of time evolution.
We have shown that the presence of  CPV in the Hamiltonian leads to
a deviation of $e^{i \ss}$ from $\pm 1$ in presence of an adiabatic
evolution iff the CPV phase also varies in space or in time.

\section{The cross terms and the action of CP}
\label{sec4} Let us examine the form of cross terms appearing in
general expression (Eq.~(\ref{prob})) for probability \bea
{\mathfrak{T}} &=& [\scx{\nu_\al}{\vartheta_1,\varphi_1,{-}}e^{ i
{{\cal D}_-}} \scx{
 \vartheta
_2,\varphi_2,{-}}{\nu_\be}
\scx{\nu_\be}{\vartheta_2,\varphi_2,{+}} e^{ -i {\cal D}_+}
\scx{\vartheta_1,\varphi_1,{+}}{\nu_\al} + {\rm{c.c.}} ]
\nonumber\\
&=& [r e^{i\ss} e^{ i {({\cal D}_- - {\cal D}_+)}} + r e^{-i \ss}
e^{ -i  {({\cal D}_- - {\cal D}_+)}} ]
\nonumber\\
&=& 2 r \cos [\ss + ({\cal D}_- - {\cal D}_+)]
\nonumber\\
&=& 2 r [ \cos \ss \cos ({\cal D}_- - {\cal D}_+) - \sin \ss \sin
({\cal D}_- - {\cal D}_+) ]~,
 \label{probcross} \eea
where  $r$ is the (real) amplitude of the cross term and $r \leq 1$.
Geometrically we can see that the amplitude $r$ will not
 change under CP. Inner product of two rays is a measure of distance
 between the two rays and this distance will not change under
 CP (reflection) operation.

 From the above expression (Eq.~(\ref{probcross})), it is
clear that for the special case when $\ss=\pi$ or $0$ i.e. when the
phase is topological in character, $\sin \ss \to 0$ and we have only
the first term ($\cos \ss$ term) contributing to the probability.
However, when the phase becomes geometric, we will have both the
terms contributing to the cross term.

Now we discuss the effect of CP transformation on the expression for
transition probability (Eq.~(\ref{prob})). CP transformation is an
anti-unitary operation  like the time-reversal operation and is
given by $\varphi_{i} \to - \varphi_i$ ($i=$1,2) \bea
{\mathfrak{T}}^{CP} &=&
[\scx{\nu_\al}{\vartheta_1,-\varphi_1,{-}}e^{ i {{\cal D}_-}} \scx{
 \vartheta
_2,-\varphi_2,{-}}{\nu_\be}
\scx{\nu_\be}{\vartheta_2,-\varphi_2,{+}} e^{ -i {\cal D}_+}
\scx{\vartheta_1,-\varphi_1,{+}}{\nu_\al} + {\rm{c.c.}} ]
\nonumber\\
&=& [r e^{-i\ss} e^{ i {({\cal D}_- - {\cal D}_+)}} + r e^{i \ss}
e^{ -i  {({\cal D}_- - {\cal D}_+)}} ]
\nonumber\\
&=& 2 r \cos [(-\ss) + ({\cal D}_- - {\cal D}_+)]
\nonumber\\
&=& 2 r [ \cos  \ss \cos ({\cal D}_- - {\cal D}_+) + \sin \ss \sin
({\cal D}_- - {\cal D}_+) ]~.
 \label{probcrosscp}
 \eea
CP transformation connects two different Hamiltonians or two
different backgrounds that are CP partners, connected by
$\varphi_{i} \to - \varphi_i$ ($i=1,2$). If we look at the form of
cross term (Eq.~(\ref{probcross})) and CP-transformed cross term
(Eq.~(\ref{probcrosscp})), we note that the cross term and hence the
transition probability will in general be different for the two CP
partners. For the special case, when  the phase is topological
($\sin \ss \to 0$) irrespective of the CP properties of the medium,
the probability is invariant under CP. The cross term will no longer
remain invariant if the presence of CPV is coupled with an adiabatic
evolution as the phase $\ss$ becomes geometric. Pairwise the two
 Hamiltonians (connected by CP transformation) are
distinguishable at the level of probability and CPV coupled with
adiabatic evolution plays a crucial role for this to happen.

We now discuss the implications of Wigner's theorem and ray space
isometries in the present context. We can draw a connection between
the above results and those rigorously
  obtained in Ref.~\cite{samray} for any quantum mechanical
  system. It was pointed out that Wigner's theorem and the isometries of ray
space can be most easily understood by using geometric ideas. The
Pancharatnam phase $\ss$ itself  is not an invariant under
isometries of the ray space, but $\cos \ss$ is. So under isometries
of the ray space, $\ss$  can be preserved  (unitary rotations) or
reversed (anti-unitary reflections). Wigner\footnote{A complete
algebraic proof of Wigner's theorem is given in
Ref.~\cite{bargmann}.} proved that any ray space isometry can be
realized on the Hilbert space of quantum mechanics by either a
unitary or anti-unitary transformation. The only measurable quantity
is the transition probability which is defined as an overlap between
rays and measures the distance between them. This result is captured
in Eqs.~(\ref{probcross}) and (\ref{probcrosscp}). In two flavor
case, when the geometric phase is restricted to be topological (CPC
case and CPV in constant density situation),
 the cross terms in the
probability depend only on $\cos \ss$ and we can not distinguish
between unitary ($\ss \to \ss$) and anti-unitary ($\ss \to -\ss$)
operations. But,
 in case the phase is geometric as in case of adiabatic evolution
coupled with CPV, we have an extra term $\propto \sin \ss$ which
will change sign under CP transformation if CP is violated.
 In our present work, we have given a geometric visualization of the
operation of CP in case of two flavor neutrino oscillations, which
 is anti-unitary. Thus, for the two state system, the \poi
sphere description is quite rich and it encompasses the full set of
possible  operations, including unitary as well as anti-unitary
operations.

\section{Conclusion}
\label{concl} We have shown that inclusion of CPV in the two flavor
neutrino Hamiltonian can  make the Pancharatnam phase geometric (and
path-dependent) from topological, in case of adiabatic evolution of
the mass eigenstates only when the CPV phases also vary with
distance.  The standard quantum mechanical treatment of
 two flavor neutrino oscillation contains Pancharatnam's phase which
  no longer remains
topological when varying CPV phases and varying background density
are taken into account.  We have given a
 neat geometric illustration
  of the fact that the two flavor oscillation probability is sensitive to the
presence of CPV phase in the Hamiltonian for the case adiabatically
varying density profile.

Concerning the observability of the geometric phase, we should note
that
 the cross term in the probability should be accessible to
 measurements. We mentioned two physical settings in Sec.~\ref{sec2} where
 the  CPV phase could appear in the Hamiltonian.

  In the first case of ordinary matter and CPV induced by NSI in propagation, we should
   note that the extrinsic CPV phases are due to the $\ep_{\al \be}^f$
    terms, which  do not  vary with distance.
    This would correspond to case (A) (if density is constant) or case (C) (if density is varying)
    described in Sec.~\ref{sec3}. In both the cases, we have shown that the phase remains topological.
A special case (R-parity violating supersymmetry) of non-standard
interactions in ordinary matter was pointed out in Ref.~\cite{he} to
show that in two flavor case, it is possible to have more than one
essential parameters (electron, proton and neutron number densities)
leading to non-zero Berry's~\cite{Berry:1984jv} phase if the cyclic
condition was also met in certain long baseline neutrino oscillation
experiments.  Note that  there are  other more exotic non-standard
ways to get flavor off-diagonal and CPV terms leading to a
modification of the dispersion relation in the neutrino Hamiltonian
for example, violation of Lorentz invariance or CPT
violation~\cite{Datta:2003}.

The second case discussed  in Sec.~\ref{sec2} was that of dense
neutrino backgrounds where CPV can appear even when SI are taken
into consideration and this was solely due to the fact that
background neutrinos in mixed state can induce CPV terms in the
Hamiltonian. Neutrinos traversing dense supernova cores in a region
outside the neutrinosphere pass through varying electron and
neutrino number densities~\cite{raffeltbook}. This could actually
correspond to case B of Sec.~\ref{sec3} if the CPV phase was also
varying with distance where the Pancharatnam prescription predicts a
geometric phase.  In case of neutrinos coming  from Supernovae
 it is impossible to detect the effects due to CPV induced geometric phases
 since the oscillations would get averaged out by the
 time the neutrinos reach the detectors on the Earth and the cross term
  will not be accessible to measurements.

If in a hypothetical situation, it was possible to realise both
 varying CPV phases and varying density, then Pancharatnam's
prescription predicts a geometric phase in the transition
probability different from $\pi,0$. The problem then reduces to
disentangling the geometric component from the total phase in the
transition probability. The dynamical phase depends on the energy
while the geometric part is independent of energy. So, by carrying
out measurements at different energy values, we can extract the
geometric part from the total phase measured in oscillation
experiments.

To conclude it turns out that with only two flavors the robustness
of the topological phase is very hard to destroy even in the
presence of CPV as long as the density and/or the CPV phases are
constant. For the two physical settings discussed above, the phase
remains topological whether CP is violated or not.  Thus, it calls
for a study of the full three flavor situation where the physics of
geometric phase can play a role~\cite{pm3f} due to the presence of
intrinsic CPV phase in the leptonic mixing matrix.

\noindent {\bf{Acknowledgments :-}} The author is deeply indebted to
Joseph Samuel and Supurna Sinha for numerous useful discussions
leading to the present work and critical comments on the manuscript.
We acknowledge Michael V. Berry for helpful email correspondence. It
is a pleasure to thank Sudhir K. Vempati for many valuable
discussions related to neutrinos and Walter Winter for helpful
discussions and comments. We thank Basudeb Dasgupta and Amol Dighe
for discussion concerning supernova neutrinos.

\providecommand{\href}[2]{#2}\begingroup\raggedright
\endgroup


\end{document}

\end{document}